\documentclass[lettersize,journal]{IEEEtran}
\usepackage{amsmath,amssymb,amsthm,mathtools,bm}
\usepackage{graphicx,cite,algorithm,algpseudocode,url,stfloats}
\allowdisplaybreaks[4]
\theoremstyle{plain}
\newtheorem{proposition}{Proposition}

\theoremstyle{remark}

\DeclareMathOperator*{\argmax}{arg\,max}

\newcommand{\norm}[1]{\left\lVert#1\right\rVert}

\newcommand{\E}{\mathbb{E}}
\begin{document}

\title{Near-Field Localization Beyond Bandwidth Limits for Large-Aperture Pinching-Antenna Systems}
\author{Hyeonho Noh,~\IEEEmembership{Member,~IEEE}%
\thanks{Hyeonho Noh is with the Department of Information and Communication Engineering, Hanbat National University, Republic of Korea (e-mail: hhnoh@hanbat.ac.kr).}}
\maketitle

\begin{abstract}
Conventional bandwidth-limited ranging resolves delay on the scale of \(c_0/(2B)\), yet a large distributed aperture can support substantially finer localization through near-field carrier-phase diversity. This letter characterizes the coherent main-lobe width of the localization likelihood for pinching-antenna systems and shows that it is governed by the spread of the direction cosines subtended by the aperture at the target. The resulting expression recovers classical far-field angular and Fresnel-range scalings and reduces to a wavelength-scale cell when the aperture subtends a large solid angle. Guided by this characterization, a hierarchical maximum-likelihood search is developed, combining noncoherent coarse localization, coherent lobe selection, and local refinement. Numerical results under scattering demonstrate submillimeter localization above a threshold signal-to-noise ratio without localization failures in the tested ensemble.
\end{abstract}


\section{Introduction}
Pinching-antenna systems have emerged as a flexible architecture in which small radiating elements are placed at selected locations along dielectric waveguides, allowing the effective aperture to be reconfigured without adding radio-frequency chains \cite{Liu26_PASSArch,Ding25_Perspective}. Because the waveguides can span several meters across rooms or corridors, the resulting aperture can be far larger than that of a conventional array at the same carrier frequency, placing the service area deep within the radiating near field.

The large and reconfigurable aperture of pinching-antenna systems has also motivated growing interest in integrated sensing and communication, with recent studies developing wideband channel and sensing models that account for guided-mode dispersion, attenuation, and frequency-dependent coupling \cite{Xiao25_OFDMPASS,Wang25_SensingPASS,Li26_PASSISAC,Wang25_PASSISAC}. Existing localization methods, however, have largely relied on the envelope of the delay response, and their resolution is therefore governed by the occupied bandwidth \cite{Noh26_DiPL}. In a large-aperture near-field deployment, the carrier phase can provide substantial additional localization information because target displacement induces distinct phase variations across the distributed radiating elements. Although carrier phase has been exploited for narrowband localization along a single waveguide \cite{Feng26_PhaseAware}, the spatial resolution enabled by this near-field phase diversity has not yet been characterized.

This letter addresses this gap in two steps. First, the carrier-phase variation induced by target displacement across a large near-field aperture is modeled, and the resulting coherent main-lobe width is derived in closed form. The expression reveals how the localization resolution depends on the aperture geometry and wavelength, recovers the classical far-field angular and Fresnel-range scalings, and approaches a wavelength-scale cell when the aperture subtends a large solid angle. Second, guided by this characterization, a hierarchical maximum-likelihood estimator is developed, combining noncoherent coarse localization, coherent lobe selection, and local refinement, with the grid spacing and search window determined from the derived resolution and the preceding-stage error. Numerical results under scattering demonstrate submillimeter localization performance above a threshold signal-to-noise ratio (SNR).

\section{Signal Model and Exact Likelihood}
\label{sec:model}
\subsection{Geometry and Guided-Wave Response}
The base station (BS) comprises $N_{\mathrm w}$ waveguides, each carrying $M$ radiating pinching antennas, together with a colocated $N_{\mathrm R}$-element receive array. The position of the $m$-th antenna on waveguide $n$ is represented by
\begin{align}
\bm\psi_m^{(n)}=\left[x_m^{(n)},y^{(n)},d_h\right]^{\mathrm T}. \label{eq:pa_position}
\end{align}
The BS utilizes $N_{\mathrm c}$ subcarriers with $f_i=f_{\mathrm c}+(i-\tfrac{N_{\mathrm c}-1}{2})\Delta f$, or equivalently $f_i=f_0+i\Delta f$ with $f_0\triangleq f_{\mathrm c}-\tfrac{N_{\mathrm c}-1}{2}\Delta f$ denoting the lowest subcarrier frequency. For a waveguide with cutoff frequency $f_{\mathrm{cut}}$, the guided-mode phase constant is
\begin{align}
\beta_g(f)=\frac{2\pi}{c_0}\sqrt{f^2-f_{\mathrm{cut}}^2}, \label{eq:beta_g}
\end{align}
which determines the frequency-dependent phase accumulated along the waveguide. The corresponding feed-to-element response reads $q_m^{(n)}(f)=\chi_m^{(n)}(f)e^{-\alpha_g(f)x_m^{(n)}}e^{-j\beta_g(f)x_m^{(n)}}$, where $\alpha_g(f)$ accounts for propagation loss and $\chi_m^{(n)}(f)$ for coupling from the guided mode to element $(n,m)$; both are assumed known from prior characterization.

\subsection{Received Signal}
During orthogonal frequency division multiplexing (OFDM) symbol $q \in \{ 1,\ldots,Q\}$, waveguide $n$ transmits a known symbol $x_q^{(n)}[i]$ on subcarrier $i$. Let $r_m^{(n),\mathrm T}(\bm p)=\norm{\bm p-\bm\psi_m^{(n)}}$ and $r_r^{\mathrm R}(\bm p)$ denote the distances from a hypothesized target position $\bm p$ to transmit element $(n,m)$ and receive element $r$, respectively. Then, the bistatic propagation delay is given by
\begin{align}
\tau_{nmr}(\bm p)=\frac{r_m^{(n),\mathrm T}(\bm p)+r_r^{\mathrm R}(\bm p)}{c_0}. \label{eq:branch_delay}
\end{align}
Defining $a_m^{(n)}[i]\triangleq\varrho(f_i)q_m^{(n)}(f_i)$, where $\varrho(f_i)$ collects the branch-independent free-space factor, the noiseless line-of-sight (LoS) response of a unit-reflectivity point target at $\bm p$ is
\begin{align}
T_q[r,i;\bm p]=\sum_{n,m}b_{nmr}(\bm p)a_m^{(n)}[i]x_q^{(n)}[i]e^{-j2\pi f_i\tau_{nmr}(\bm p)}, \label{eq:exact_template}
\end{align}
where $b_{nmr}(\bm p)=[r_m^{(n),\mathrm T}(\bm p)r_r^{\mathrm R}(\bm p)]^{-1}$ denotes the position-dependent amplitude of branch $(n,m,r)$. Let $\mathbf T_q(\bm p)$ collect $T_q[r,i;\bm p]$ over the receive elements and subcarriers. The received signal is modeled as $\mathbf Y_q=\gamma\mathbf T_q(\bm p^\star)+\mathbf N_q$, where $\bm p^\star$ is the true target position, $\gamma$ is its unknown complex reflectivity, and $\mathbf N_q$ is additive white Gaussian noise.

\subsection{Efficient Likelihood Evaluation}

For a hypothesized target position $\bm p$, maximizing the Gaussian likelihood over the unknown reflectivity $\gamma$ yields
\begin{align}
\widehat{\bm p}=\argmax_{\bm p}\Lambda(\bm p), \quad
\Lambda(\bm p)=\frac{\left|\sum_{q=1}^{Q}\langle\mathbf T_q(\bm p),\mathbf Y_q\rangle_\text{F}\right|^2}{\sum_{q=1}^{Q}\norm{\mathbf T_q(\bm p)}_\text{F}^2}.
\label{eq:profiled_ml}
\end{align}
Direct evaluation of \eqref{eq:profiled_ml} requires forming $\mathbf T_q(\bm p)$ for every hypothesis. Since the transmit symbols and guided-wave responses are independent of $\bm p$, their contributions can be removed beforehand. For each transmit--receive element pair $(n,m,r)$, define
\begin{align}
\zeta_{nmr,q}[i]&=Y_q[r,i]\,x_q^{(n)*}[i]\,a_m^{(n)*}[i], \label{eq:deembedded_spectrum}\\
G_{nmr}(\tau)&=\sum_{i=0}^{N_{\mathrm c}-1}\left(\sum_{q=1}^{Q}\zeta_{nmr,q}[i]\right)e^{j2\pi i\Delta f\tau}. \label{eq:baseband_profile}
\end{align}
The profiles $G_{nmr}(\tau)$ are computed once using zero-padded fast Fourier transforms (FFTs). Using $f_i=f_0+i\Delta f$, the coherent matched-filter term becomes
\begin{align}
S(\bm p)=\sum_{n=1}^{N_{\mathrm w}}\sum_{m=1}^{M}\sum_{r=1}^{N_{\mathrm R}}
b_{nmr}(\bm p)e^{j2\pi f_0\tau_{nmr}(\bm p)}
G_{nmr}\!\left(\tau_{nmr}(\bm p)\right),
\label{eq:coherent_score}
\end{align}
Note that since $\{G_{nmr}\}$ depend only on the received data and the characterized hardware, they are computed once and reused for all hypotheses, reducing each likelihood evaluation to one delay interpolation per transmit--receive element pair. The carrier-phase term is evaluated analytically because, at millimeter-wave frequencies, it can vary by more than one cycle between adjacent delay samples and is therefore not reliably captured by interpolation.

\section{Near-Field Ambiguity and Hierarchical Search}
\label{sec:ambiguity}
\subsection{Coherent and Noncoherent Search Statistics}

For each radiating element $(n,m)$, define its coherent receive-array contribution as
\begin{align}
c_{nm}(\bm p)=\sum_{r=1}^{N_{\mathrm R}} b_{nmr}(\bm p)e^{j2\pi f_0\tau_{nmr}(\bm p)}G_{nmr}\!\left(\tau_{nmr}(\bm p)\right).
\label{eq:element_score}
\end{align}
This gives the coherent and noncoherent statistics
\begin{align}
\Lambda(\bm p)&=\frac{\left|\sum_{n=1}^{N_{\mathrm w}}\sum_{m=1}^{M}c_{nm}(\bm p)\right|^2}{E(\bm p)}, \label{eq:coherent_stat}\\
\Lambda_{\mathrm{nc}}(\bm p)&=\frac{\left(\sum_{n=1}^{N_{\mathrm w}}\sum_{m=1}^{M}\left|c_{nm}(\bm p)\right|\right)^2}{E(\bm p)}, \label{eq:noncoherent_stat}
\end{align}
where $E(\bm p)=\sum_{q=1}^{Q}\norm{\mathbf T_q(\bm p)}^2$ is the template energy for the known transmitted symbols and hardware responses. By the triangle inequality, $\Lambda(\bm p)\leq\Lambda_{\mathrm{nc}}(\bm p)$ for all $\bm p$.

\subsection{Coherent Main-Lobe Width}

Let $\bm u_{nm}(\bm p)$ denote the unit vector from $\bm p$ to radiating element $(n,m)$. For a displacement of magnitude $\Delta$ along a unit direction $\hat{\bm e}$, i.e., $\bm p'=\bm p+\Delta\hat{\bm e}$, the first-order delay variation is
\begin{align}
\tau_{nmr}(\bm p')-\tau_{nmr}(\bm p)
\simeq-\frac{\Delta}{c_0}\left(\bm u_{nm}(\bm p)\cdot\hat{\bm e}+\bm v_r(\bm p)\cdot\hat{\bm e}\right),
\label{eq:delay_perturbation}
\end{align}
where $\bm v_r(\bm p)$ is the unit vector from $\bm p$ to receive element $r$. Since the receive array is compact relative to the transmit aperture, $\bm v_r(\bm p)\cdot\hat{\bm e}$ varies little across $r$ and contributes approximately a common phase. The normalized ambiguity between $\bm p$ and $\bm p'$ is therefore
\begin{align}
\mathcal A(\Delta)=
\frac{\left|\sum_{q=1}^{Q}\left\langle\mathbf T_q(\bm p'),\mathbf T_q(\bm p)\right\rangle_\text{F}\right|^2}
{\left(\sum_{q=1}^{Q}\norm{\mathbf T_q(\bm p)}_\text{F}^2\right)
 \left(\sum_{q=1}^{Q}\norm{\mathbf T_q(\bm p')}_\text{F}^2\right)},
\label{eq:ambiguity}
\end{align}
which, under the carrier-phase approximation, reduces to
\begin{align}
\mathcal A(\Delta)\simeq
\left|
\E_{nm}\!\left[
e^{j\frac{2\pi}{\lambda}\Delta\,\bm u_{nm}(\bm p)\cdot\hat{\bm e}}
\right]
\right|^2.
\label{eq:ambiguity_cf}
\end{align}
where $\E_{nm}[\cdot]$ denotes averaging over the radiating elements. Thus, the coherent main lobe is governed by the distribution of aperture direction cosines rather than by the signal bandwidth. Let $\sigma_{\hat{\bm e}}$ denote the standard deviation, over the radiating elements, of the direction cosine $\bm u_{nm}(\bm p)\cdot\hat{\bm e}$. For small $\Delta$, \eqref{eq:ambiguity_cf} is approximated by $\mathcal A(\Delta)\simeq \exp[
-(\frac{2\pi\sigma_{\hat{\bm e}}\Delta} \lambda)^2 ]$.
Setting $\mathcal A(\Delta)=1/2$ gives the full width between the two half-power points.

\begin{proposition}[Coherent main-lobe width]
\label{prop:cell}
The full width between the half-power points of the coherent main lobe along $\hat{\bm e}$ is
\begin{align}
\delta_{\mathrm{coh}}(\hat{\bm e})
=\frac{\lambda}{\pi}\frac{\sqrt{\ln 2}}{\sigma_{\hat{\bm e}}}.
\label{eq:cell_width}
\end{align}
\end{proposition}

Equation \eqref{eq:cell_width} recovers the familiar far-field scalings. For a broadside target at range $r_0$ from an aperture of extent $D$, a transverse displacement gives $\sigma_{\hat{\bm e}}\simeq D/(\sqrt{12}r_0)$ and hence
\begin{align}
\delta_{\mathrm{coh}}
\simeq\frac{\sqrt{12\ln 2}}{\pi}\frac{\lambda r_0}{D}
\approx0.92\,\frac{\lambda r_0}{D},
\label{eq:crossrange_limit}
\end{align}
consistent with the classical angular-resolution scaling. Along the line of sight, $\sigma_{\hat{\bm e}}=O(D^2/r_0^2)$, giving $\delta_{\mathrm{coh}}=O(\lambda r_0^2/D^2)$, i.e., the Fresnel range scaling. In the large-solid-angle regime relevant to pinching-antenna systems, the aperture is comparable to the target range and $\sigma_{\hat{\bm e}}=O(1)$, so
\begin{align}
\delta_{\mathrm{coh}}=O(\lambda),
\label{eq:pass_cell}
\end{align}
with anisotropy determined by the aperture geometry. In contrast, removing the relative phase across radiating elements yields the noncoherent scale $\delta_{\mathrm{nc}}\sim c_0/(2B)$.

\subsection{Hierarchical Estimation}

The separation between $\delta_{\mathrm{nc}}$ and $\delta_{\mathrm{coh}}$ motivates a hierarchical search. Let $\mathcal B(\bm c,\rho)=\{\bm p:\norm{\bm p-\bm c}_\infty\le\rho\}$ denote the square region of half-width $\rho$ centered at $\bm c$, let $\mathcal G(\mathcal C,\delta)$ denote the uniform grid of spacing $\delta$ covering a region $\mathcal C$, and let $\mathcal P$ denote the service area. The noncoherent statistic $\Lambda_{\mathrm{nc}}$ is first evaluated on $\mathcal G(\mathcal P,\delta_0)$ to obtain a coarse estimate $\bm p^{(1)}$, and then on $\mathcal G(\mathcal B(\bm p^{(1)},\rho_1),\delta_1)$, with $\rho_1$ covering the first-stage grid cell, to obtain $\bm p^{(2)}$. The coherent statistic $\Lambda$ is then evaluated on $\mathcal G(\mathcal B(\bm p^{(2)},R),\delta)$ to obtain $\bm p^{(3)}$, which initializes a local refinement yielding the final estimate $\widehat{\bm p}$. Algorithm~\ref{alg:est} summarizes the procedure.

The grid spacings are selected according to the corresponding resolution scales. The noncoherent spacings $\delta_0$ and $\delta_1$ are chosen on the bandwidth-limited scale, with $\delta_1<\delta_0$ and $\delta_1\le\delta_{\mathrm{nc}}/2$. For the coherent search, Proposition~\ref{prop:cell} gives a direction-dependent main-lobe width; for a Cartesian grid, the spacing is chosen as
\begin{align}
\delta\le\delta_{\max}\triangleq\frac{1}{2}\min_j\delta_{\mathrm{coh}}(\bm e_j),
\label{eq:coh_grid}
\end{align}
where $\bm e_j$ denotes the $j$th coordinate direction. The half-width $R$ is selected to contain the error of the preceding noncoherent estimate with probability at least $1-\varepsilon$. Defining $E_2=\norm{\bm p^{(2)}-\bm p^\star}_\infty$, this requires
\begin{align}
R\ge F_{E_2}^{-1}(1-\varepsilon),
\label{eq:window_rule}
\end{align}
where $F_{E_2}^{-1}$ denotes the quantile function of $E_2$.

Starting from $\bm p^{(3)}$, the final stage performs projected gradient ascent over $\mathcal N=\{\bm p:\norm{\bm p-\bm p^{(3)}}_\infty\le h\}$, where $h$ is chosen smaller than the selected coherent main lobe:
\begin{align}
\bm p_{k+1}=\Pi_{\mathcal N}\!\left(\bm p_k+\mu_k\nabla\Lambda(\bm p_k)\right),\qquad \bm p_0=\bm p^{(3)}.
\label{eq:ascent}
\end{align}
Here, $\Pi_{\mathcal N}$ denotes projection onto $\mathcal N$, and $\mu_k$ is selected by backtracking to ensure an increase in $\Lambda$. The gradient can be evaluated analytically from the derivatives of $\tau_{nmr}(\bm p)$, $b_{nmr}(\bm p)$, $G_{nmr}(\tau)$, and $E(\bm p)$. The projection confines the refinement to the selected lobe, while the projected gradient ascent removes the residual grid-quantization error.

\begin{algorithm}[t]
\caption{Hierarchical near-field localization}
\label{alg:est}
\begin{algorithmic}[1]
\Require data $\{\mathbf Y_q\}$, known $\{x_q^{(n)}[i]\}$ and $\{a_m^{(n)}[i]\}$, tabulated $E(\cdot)$, window $R$ from \eqref{eq:window_rule}, spacing $\delta\le\delta_{\max}$
\State $G_{nmr}\gets$ zero-padded FFT of $\sum_q\zeta_{nmr,q}$ for every branch \label{ln:fft}
\State $\bm p^{(1)}\gets\argmax_{\bm p\in\mathcal G(\mathcal P,\delta_0)}\Lambda_{\mathrm{nc}}(\bm p)$ \label{ln:scan}
\State $\bm p^{(2)}\gets\argmax_{\bm p\in\mathcal G(\mathcal B(\bm p^{(1)},\rho_1),\delta_1)}\Lambda_{\mathrm{nc}}(\bm p)$ \label{ln:refine2}
\State $\bm p^{(3)}\gets\argmax_{\bm p\in\mathcal G(\mathcal B(\bm p^{(2)},R),\delta)}\Lambda(\bm p)$ \label{ln:coh}
\State $\widehat{\bm p}\gets$ projected gradient ascent of $\Lambda$ from $\bm p^{(3)}$ \label{ln:asc}
\State \Return $\widehat{\bm p}$
\end{algorithmic}
\end{algorithm}

\section{Numerical Results}
\label{sec:results}
\subsection{Setup}

The geometry follows \cite{Xu25_GPASSTutorial}, with the waveguide and coupling responses modeled as in \cite{Xiao25_OFDMPASS}. The carrier frequency is $28$ GHz, the bandwidth is $2$ GHz over $512$ subcarriers, and the waveguide cutoff frequency is $27.3$ GHz. Eight $20$-m waveguides, each carrying eight radiating elements, are deployed at a height of $5$ m, forming an aperture of approximately $17.5$ m by $8.8$ m, together with a $16$-element receive array. Targets are drawn uniformly over the central $18$ m by $8$ m region of a $20$ m by $10$ m service area, and each waveguide transmits independent unit-power quadrature amplitude modulation symbols. Unless stated otherwise, the scattered component is generated according to a Rician model with Rician $K$-factor $K_{\mathrm R}=10$ dB. Accuracy is reported as the root-mean-square error (RMSE).

Five baselines are evaluated on the same data. Two replace the distributed aperture by a conventional array carrying the same number of elements on a half-wavelength grid, deployed either at the feed next to the receive array \cite{Feng26_PhaseAware} or at the center of the service area \cite{Jiang25_BCRB}. The remaining three isolate the stages of the proposed estimator: the noncoherent statistic \eqref{eq:noncoherent_stat} alone; the same estimate followed by local refinement; and the coherent grid without the final refinement. All schemes share the transmit power, the target ensemble, and the noise and scattering realizations, and each determines its own grid spacing and search window from the same rules. The Cram\'er--Rao bound (CRB) of the LoS model is included as a reference.

\subsection{Results}
\begin{figure}[t]
\centering
\includegraphics[width=\columnwidth]{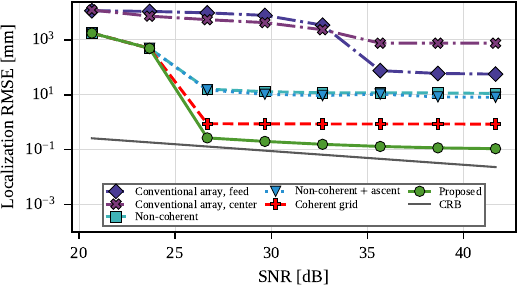}
\vspace{-20pt}
\caption{Localization RMSE versus SNR.}
\label{fig:snr}
\end{figure}
Fig.~\ref{fig:snr} compares the localization error as a function of SNR. The proposed estimator exhibits a clear threshold near $27$ dB, beyond which its RMSE approaches the CRB and gradually decreases to the submillimeter level. The noncoherent estimator also improves beyond this threshold but remains more than an order of magnitude less accurate, showing that delay information alone does not exploit the spatial resolution provided by the distributed aperture. Local refinement initialized from the noncoherent estimate gives almost no additional gain, whereas the coherent grid reduces the error to the millimeter level by resolving the correct likelihood lobe. Once this lobe is identified, the final refinement provides a further three- to sevenfold improvement. The conventional arrays require substantially higher SNR to enter the same operating regime and remain markedly less accurate over the considered range. This confirms the benefit of the large distributed aperture.

\begin{figure}[t]
\centering
\includegraphics[width=\columnwidth]{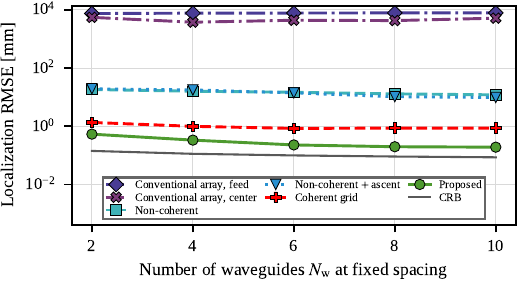}
\vspace{-20pt}
\caption{Localization RMSE versus the number of waveguides at a fixed spacing.}
\label{fig:sweep_s}
\vspace{-8pt}
\end{figure}
\begin{figure}[t]
\centering
\includegraphics[width=\columnwidth]{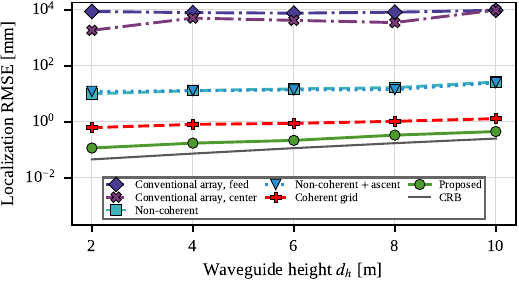}
\vspace{-20pt}
\caption{Localization RMSE versus the waveguide height.}
\label{fig:sweep_h}
\vspace{-8pt}
\end{figure}
\begin{figure}[t]
\centering
\includegraphics[width=\columnwidth]{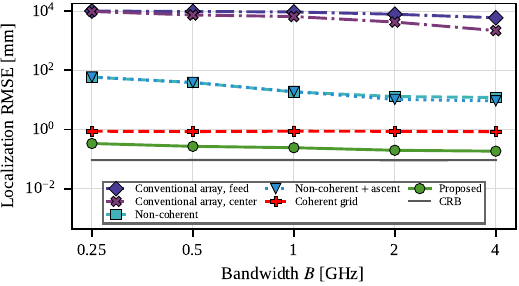}
\vspace{-20pt}
\caption{Localization RMSE versus the bandwidth.}
\label{fig:sweep_b}
\vspace{-9pt}
\end{figure}
As shown in Fig.~\ref{fig:sweep_s}, increasing the number of waveguides at a fixed spacing enlarges the distributed aperture and progressively improves the proposed estimator along with the CRB. This trend aligns with \eqref{eq:cell_width}, since the larger aperture increases $\sigma_{\hat{\bm e}}$ and narrows the coherent main lobe. In contrast, the noncoherent estimators are nearly insensitive to the aperture size because their resolution remains bandwidth limited. The coherent grid alone also exhibits only a modest improvement, indicating that lobe selection by itself does not fully exploit the increasingly narrow coherent peak. The substantial gain appears only after the local refinement, which converts the aperture-induced narrowing of the likelihood into improved positioning accuracy. 

Fig.~\ref{fig:sweep_h} illustrates the localization RMSE with respect to the waveguide height. Increasing the height reduces the angular diversity provided by the distributed aperture, as the direction cosines across the radiating elements become more concentrated. Consequently, $\sigma_{\hat{\bm e}}$ decreases and the coherent main lobe in \eqref{eq:cell_width} becomes wider. The increased propagation distance further reduces the received SNR, leading to a corresponding increase in the CRB. The proposed estimator closely follows this trend, indicating that its performance degradation is mainly dictated by the reduced geometric information rather than by the search procedure. In contrast, the coherent-grid estimator suffers from an additional quantization loss because the grid spacing increases with the coherent main-lobe width. The subsequent local refinement largely removes this loss, enabling the proposed estimator to effectively exploit the localization information available from the aperture geometry.

The impact of bandwidth is reported in Fig.~\ref{fig:sweep_b}. As the bandwidth decreases, the noncoherent resolution deteriorates according to $\delta_{\mathrm{nc}}\sim c_0/(2B)$, whereas the coherent CRB remains nearly unchanged because the carrier-phase information is primarily governed by the distributed-aperture geometry. Since the proposed estimator relies on the noncoherent stages to initialize the subsequent coherent search, its RMSE also increases as the bandwidth decreases. Nevertheless, the proposed estimator maintains a clear accuracy advantage over the considered baselines through the coherent grid search and subsequent projected gradient ascent. This demonstrates that the hierarchical search retains most of the coherent localization gain even when the delay resolution available to the initial search becomes coarse.

\section{Conclusion}
This letter investigates near-field localization with large-aperture pinching-antenna systems. The coherent localization resolution is characterized and contrasted with the bandwidth-limited noncoherent resolution, showing that the coherent main-lobe width is governed by the aperture direction-cosine spread and can achieve wavelength-scale resolution beyond the bandwidth-limited noncoherent response. Based on these characteristics, a hierarchical localization algorithm is developed by combining noncoherent coarse search, coherent lobe selection, and local refinement. Numerical results demonstrate that the proposed method effectively exploits the large distributed aperture and achieves submillimeter localization accuracy.

\bibliographystyle{IEEEtran}
\bibliography{refs}
\end{document}